\documentclass[%
 reprint,
amsmath,
amssymb,
prl,
floatfix,
]{revtex4-2}

\usepackage{graphicx}
\usepackage{dcolumn}
\usepackage{bm}
\usepackage{hyperref}
\usepackage{cleveref}
\usepackage{subcaption}
\usepackage{caption, xcolor}
\usepackage{ragged2e}
\usepackage{ulem, adjustbox}
\usepackage{changes}
\definechangesauthor[name={smc}, color=orange]{smc}
\definechangesauthor[name={jiantang}, color=red]{tj}
\definechangesauthor[name={shihanzhao}, color=blue]{zsh}

\newcommand{\pmu}{\mathcal{P}_{\mu}}
\newcommand{\Nr}{N_{\mathrm{M}} (\theta_r)}
\newcommand{\kpi}{{$K\texttt{-}\pi$ ratio}}

\begin{document}

\preprint{APS/123-QED}

\title{Cosmic Ray Muon Polarization to Facilitate Atmospheric Neutrino Physics}

\author{Ming-Chen Sun}
\author{Shihan Zhao}
\author{Rui-Xuan Gao}
\author{He-Sheng Liu}
\author{Ai-Yu Bai}
\author{Jian Tang}%
 \email{tangjian5@mail.sysu.edu.cn}
\affiliation{School of Physics, Sun Yat-sen University, Guangzhou, 510275, China}
\affiliation{Platform for Muon Science and Technology, Sun Yat-sen University, Guangzhou 510275, China}

\date{\today}

\begin{abstract}
Atmospheric neutrinos (ATNs) offer a paradigm for understanding neutrino properties, while it is critical to quantify uncertainties in flux modeling.
Since ATNs are produced simultaneously with cosmic ray muons, precision measurements of cosmic ray muons, including arrival direction, energy spectra, and spin polarization, will help reduce ATN production uncertainties and facilitate atmospheric neutrino physics.
This letter proposes using an array strategy to measure the spin polarization of cosmic ray muons, thereby strengthening the emergent synergies between cosmic ray and atmospheric neutrino physics.
Constraints on long-standing atmospheric neutrino flux uncertainties at the percentage level in a few-GeV energy range are achievable within one year using a $\mathcal{O}(10)~\text{m}^2$ array of Cosmic-Ray muon Spin polarization detectoRs (CRmuSRs).
With the resulting reduction in flux uncertainties, oscillation analysis of atmospheric neutrinos in a liquid scintillator detector with an exposure of 1500$\text{kt}\cdot\text{yr}$ will break the octant degeneracy and achieve the precision measurement of $\theta_{23}$ with the uncertainty smaller than $5^\circ$ at 3$\sigma$ confidence level irrespective of the mass ordering.
\end{abstract}

\maketitle

\noindent\textbf{Introduction.} Cosmic rays (CRs) consist of primary particles originating from space and secondary particles produced by their interactions with the Earth's atmosphere~\cite{Gaisser:2016uoy, cosmicRayReview}.
CRs are present in almost all particle physics and nuclear physics experiments, either as a signal or a background source.
Satellite experiments~\cite{aguilar2021alpha,Adriani:2018ktz,DAMPE:2017fbg,Fermi-LAT:2017bpc,HESS:2008ibn,PAMELA:2013vxg} in Earth's orbit directly measure the energy spectrum and composition of primary cosmic rays below $100~\mathrm{TeV}$.
Ground-based CR observation experiments~\cite{Salamida:2023fmk,Bezyazeekov:2018yjw,Schellart:2013bba} reconstruct extensive air showers to search for ultra-high-energy CR events and reveal the acceleration mechanisms of CRs.
Accelerator and collider experiments~\cite{LHCVeto,CometStatus,Mu2e,bai2024conceptual} treat CRs as a background and reject them by veto detectors to improve the experimental sensitivity.
For deep underground experiments, such as dark matter searches~\cite{ADMX:2018gho,LUX-ZEPLIN,XENON:2024ijk,PandaX:2024muv} and neutrino experiments~\cite{Super-Kamiokande:2002weg,DUNE:2020lwj,IceCube:2014stg}, sensitivity is hindered by the ATNs and high-energy CR muon events~\cite{PhysRevD.9.1389,COHERENT:2017ipa,COHERENT:2020iec,COHERENT:2021xmm,Tang:2023xub,Tang:2024prl,Liao:2025hcs}.
Therefore, it is essential to model cosmic rays with high precision to facilitate the existing and future particle physics experiments.

On the other hand, atmospheric neutrinos offer a unique opportunity to study neutrino properties.
Neutrino oscillations reveal the non-trivial flavor structure of the lepton sector, motivating a precise determination of the neutrino mixing parameters.
However, determination of $\theta_{23}$ still suffers from octant ambiguity, while $\delta_{\text{CP}}$ strongly depends on the neutrino mass ordering~\cite{Esteban:2024eli}. Reducing flux model uncertainties could help resolve these ambiguities.
Neutrino oscillation parameters have been measured using reactor neutrinos~\cite{DayaBay:2012fng,RENO:2012mkc,DoubleChooz:2011ymz}, accelerator neutrinos~\cite{T2K:2011ypd,NOvA:2019cyt}, and atmospheric neutrinos~\cite{Super-Kamiokande:2023ahc,KM3NeT:2021ozk,Hill:2025vmu}.
Due to the wide energy range and long travel distance, the ATNs are widely used to investigate $\theta_{23}$ and $\Delta m_{32}^2$.
Therefore, various ATN experiments are under construction or in the upgrade phase~\cite{JUNO:2015zny,JUNO:2021vlw,Jena:2024ycp,Nakamura:2003hk,IceCube-Gen2:2020qha}, requiring precise primary ATN flux as an essential input.
Tremendous effort has been devoted to constructing low-threshold yet high-resolution neutrino detectors to extract more precise information from oscillated atmospheric neutrinos.
In contrast, the current uncertainties of the low-energy ATNs remain ambiguous; however, rich physics is embedded in the energy range smaller than 1 GeV ~\cite{Honda:2015fha, Barr:2006it, Battistoni:2002ew, Evans:2016obt}.
In the near future, the mixing parameters will continue to be investigated by experiments such as JUNO~\cite{JUNO:2021vlw}, Hyper-Kamiokande~\cite{Nakamura:2003hk}, and DUNE~\cite{DUNE:2020lwj}. 
\begin{figure}[htpb]
    \centering
    \includegraphics[width = \linewidth]{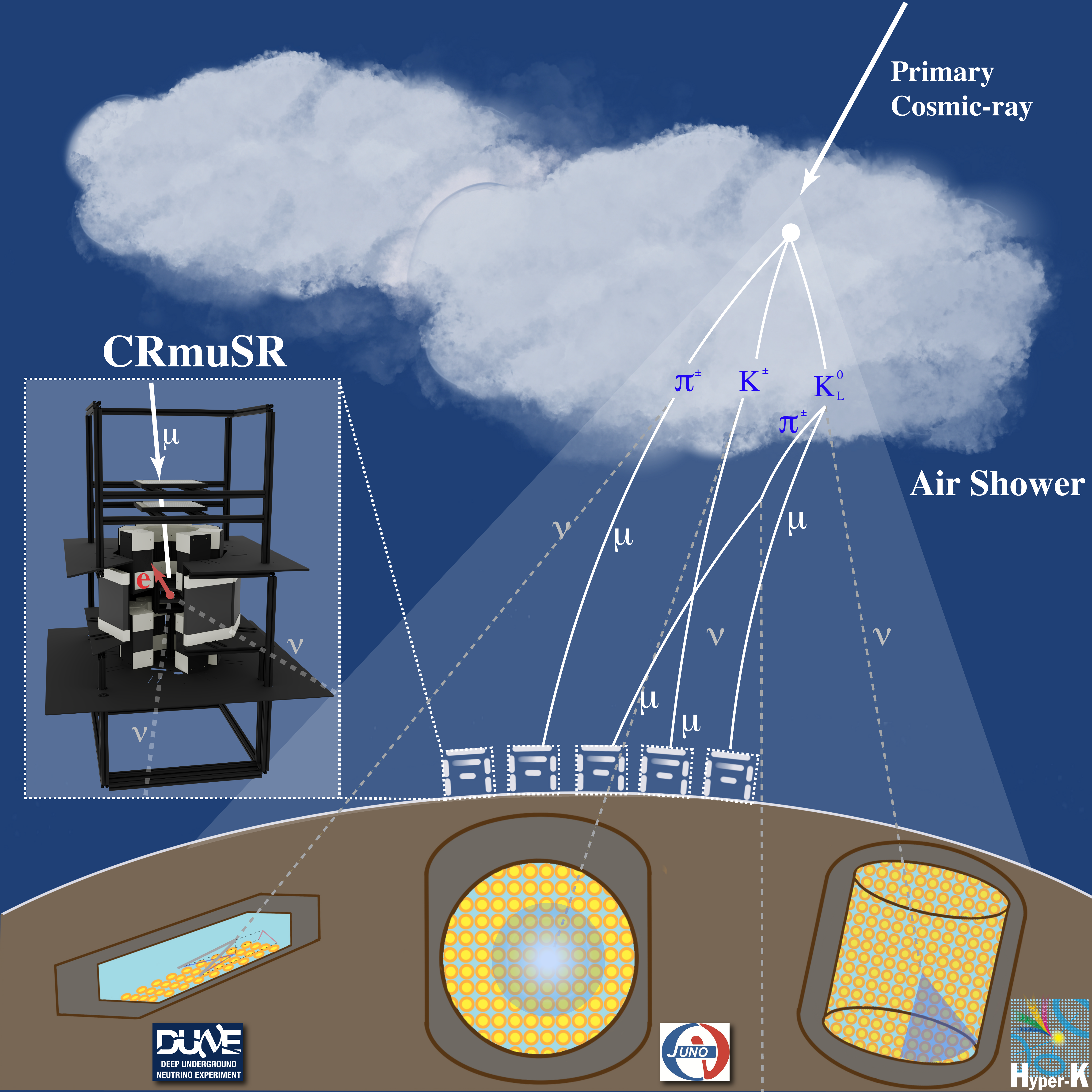}
    \caption{\justifying An artistic representation of the measurement of cosmic-ray muon polarization by CRmuSR. The CRmuSR array (middle left) provides an opportunity to constrain the production processes of ATNs based on experimental measurements of the corresponding muons' polarization.}
    \label{fig:CRmuSR_ART}
\end{figure}

As the dominant component of CRs at the sea level, cosmic-ray muons are intimately connected to ATNs~\cite{barr1988ratio, barr1989flux}.
Cosmic ray muon experiments not only provide critical data to refine the CR modeling but also offer a unique opportunity to probe the ATN production mechanisms.
In this article, we demonstrate the feasibility of a new experimental strategy to measure cosmic-ray muon polarization, aiming to provide complementary constraints on ATN production process, as illustrated in~\cref{fig:CRmuSR_ART}.
To achieve this goal, we propose measuring cosmic ray muon spin polarization using both muon momentum and the spatial distribution of Michel electrons.
We will present a dedicated detector system designed to reconstruct muon polarization, meeting physics requirements with a modular design that facilitates the deployment of large-scale arrays and future integration with neutrino experiments.
Furthermore, the proposed method provides an independent constraint on low-energy ATN production, offering a complementary approach to traditional accelerator-based measurements.

\noindent\textbf{ATN flux and cosmic-ray muon polarization.} 
In the measurement of neutrino oscillation parameters, a reliable estimation of the atmospheric neutrino flux is essential, as it sets the foundation for interpreting experimental observations.
According to the study by Gaisser and Honda~\cite{gaisser2002flux}, the ATN flux can be represented as
\begin{equation}
    \phi_{\nu_i} = \phi_p \otimes R_p \otimes Y_{p \to \nu_i} + \sum_A
    \left\{
        \phi_A \otimes R_A \otimes Y_{A \to \nu_i}
    \right\}, 
\end{equation}
where $i$ is the neutrino flavor, $\phi_p$ represents the primary cosmic-ray flux, $R_p$ is the geomagnetic field selection effect, and $Y(A \to \nu_i)$ is the neutrino yield per primary particle.
Note that the free and bound nucleons are treated separately due to their different rigidity.
As the mixing parameters are derived by fitting the ATN flux $\phi_{\nu_i}$ model to reconstructed experimental data, the uncertainty in $\phi_{\nu_i}$ will directly impact the precision measurement of the neutrino mixing parameters. 

\begin{figure}[htpb]
    \centering
    \includegraphics[width = \linewidth]{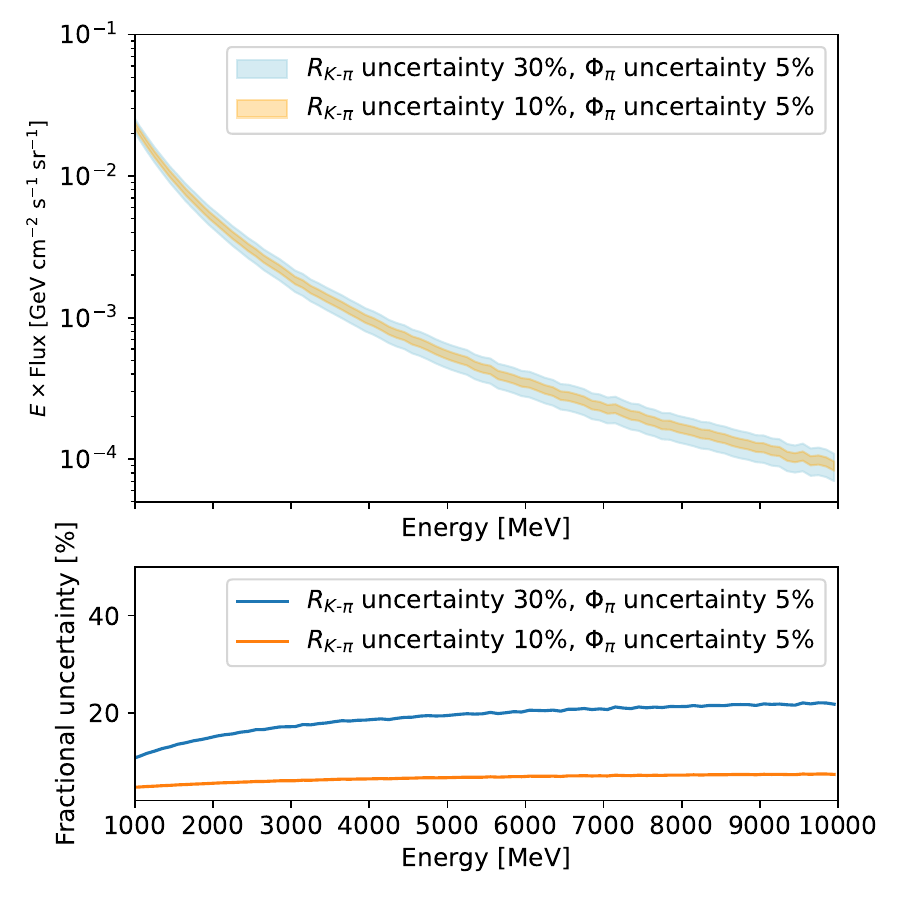}
    \caption{\justifying The upper panel shows the atmospheric neutrino flux as a function of energy, while the lower panel presents the corresponding fractional uncertainty. The shaded bands illustrate the impact of varying assumptions on the uncertainty of the {\kpi}, with the pion production uncertainty fixed at $5\%$. The blue band corresponds to an assumed $30\%$ uncertainty in {\kpi}, and the orange band to a $10\%$ uncertainty.}
    \label{fig:atn_uncertainty}
\end{figure}

There are three primary sources of uncertainty in the current ATN flux modeling.
One is the uncertainty in the energy spectrum and composition of primary cosmic rays, associated with $\phi_p$ and $\phi_A$, which originate from several physical effects, such as solar modulation.
Another is the uncertainty associated with the geomagnetic field, which primarily arises from the temporal evolution of the geomagnetic field, the limited resolution of the field models, the directional dependence of particle deflection, and the coupling with solar activity, particularly through $R_p$ and $R_A$.
Lastly, there is the uncertainty of $Y_{p \to \nu_i}$, which arises from atmospheric density fluctuations caused by temperature changes and the uncertainty of the strong interaction cross section between primary cosmic rays and atmospheric nucleons.
Primary cosmic-ray measurements, such as AMS-02~\cite{AMSPrimaryParticle} and PAMELA~\cite{PAMELA:2013vxg}, have significantly reduced the uncertainty in $\phi_p$.
Meanwhile, the uncertainty of $R_p(R_A)$ has been significantly suppressed by employing high-precision geomagnetic field models such as IGRF~\cite{alken2021international}, which ensure reliable estimation of geomagnetic cutoff and directional effects.
On the contrary, the precise determination of $Y_{p \to \nu_i}$ remains a challenge.
Accelerator-based experiments measure the cross sections of strong interactions within a limited phase space, where the pion momentum $p_\pi < 4~\mathrm{GeV}$ has not been covered yet~\cite{gaisser2002flux}.
It is highly nontrivial to quantify the dominant contribution from $Y_{p \to \nu_i}$.
Since ATNs are mainly produced in the decays of $\pi^\pm$, $K^\pm$, $K_L^0$, and $\mu^\pm$, their production channels can be schematically summarized as
\begin{equation}
    \begin{aligned}
        & \pi^+ \to \mu^+ + \nu_\mu~, && \pi^- \to \mu^- + \bar{\nu}_\mu~, \\
        & K^+ \to \mu^+ + \nu_\mu~,  && K^- \to \mu^- + \bar{\nu}_\mu~, \\
        & K_L^0 \to \mu^+ + \pi^- + \nu_\mu~, && K_L^0 \to \mu^- + \pi^+ + \bar{\nu}_\mu~, \\
        & \mu^+ \to e^+ + \nu_e + \bar{\nu}_\mu~, && \mu^- \to e^- + \bar{\nu}_e + \nu_\mu~.
    \end{aligned}
\label{eq:muon_production_mode}
\end{equation}
In these processes, ATNs are predominantly produced by pion, kaon, and muon decays, while the cosmic-ray muons themselves mainly originate from pions and kaons. 
Therefore, the total atmospheric neutrino flux can be expressed as the sum of component fluxes associated with different parent particles. 
Under this interpretation, the absolute flux sets the overall normalization, and the relative contributions of kaon and pion production (\kpi) are essential for characterizing atmospheric neutrino production.
To investigate how the uncertainty of {\kpi} propagates to the flux, we developed the Mustard-based Air Shower simulation toolkit (MusAirS). 
MusAirS is developed based on the Mustard framework~\cite{MustardFramework}, which is used to simulate modern high-energy physics experiments, such as the Muonium-to-Antimuonium Conversion Experiment (MACE)~\cite{bai2024conceptual}.
MusAirS models extensive air showers initiated by the interactions of primary cosmic rays with the Earth’s atmosphere. The input primary spectrum and composition are based on AMS-02 measurements~\cite{AMSPrimaryParticle}, where only protons and $\alpha$ particles are included since they account for about $99.9\%$ of the total flux. 
The atmosphere is described by the International Standard Atmosphere (ISA) up to an altitude of $85~\mathrm{km}$, which is sufficient given that most showers relevant to ATN and muon production occur around $20~\mathrm{km}$ above sea level. 
A uniform geomagnetic field is implemented according to the local conditions near Zhaoqing~\cite{ZhaoqingGMF2021} (close to our laboratory in Guangzhou, China). By tracking both parent and grandparent particles, MusAirS allows us to disentangle neutrino production channels.
Accordingly, the ATN flux can be modeled as a superposition of channel-dependent contributions:
\begin{equation}
    \begin{aligned}
        \Phi_{\text{ATN}}(E) =& \, \Phi_{\pi\to\text{ATN}}(E) + \Phi_{K\to\text{ATN}}(E)
        + \Phi_{\pi\to\mu\to\text{ATN}}(E) \\
        &+\Phi_{K\to\mu\to\text{ATN}}(E) + \Phi_{\rm other}(E).
    \end{aligned}
\end{equation}
The contribution from each parent particle is represented by a spectral template $s_i(E)$, which is derived from simulation and normalized using the corresponding parent flux. 
By taking $\Phi_\pi(E)$ as the reference and introducing the {\kpi} $R_{K\text{-}\pi} \equiv \Phi_K / \Phi_\pi$, the total flux can be expressed as
\begin{equation}
    \label{eq:atn_flux_component}
    \begin{aligned}
        \Phi_{\text{ATN}}(E)
        = &\Phi_{\pi}(E)\Big[
        s_{\pi\to\text{ATN}}(E) + R_{K \text{-}\pi} \cdot s_{K\to\text{ATN}}(E) \\
        & + s_{\pi\to\mu\to\text{ATN}}(E) + R_{K \text{-}\pi} \cdot s_{K\to\mu\to\text{ATN}}(E)
        \Big] \\
        & + \Phi_{\rm other}(E).
    \end{aligned}
\end{equation}
In practice, we generate the templates $s_i(E)$ with MusAirS and construct an ATN flux to investigate how the flux uncertainty varies under the uncertainty of {\kpi} and $\Phi_\pi (E)$ variances.
In the simulation shown in~\cref{fig:atn_uncertainty}, both the uncertainty of $\Phi_\pi(E)$ and the $R_{K\text{-}\pi}$ are concerned.
Consequently, precise control of $R_{K\text{-}\pi}$ and $\Phi_{\pi}(E)$ can substantially constrain the uncertainty of atmospheric neutrino production. 
In particular, an improved determination of $R_{K\text{-}\pi}$ reduces the flux uncertainty from the $\sim 10\%$ level to only a few percent.

To investigate the impact of improving the uncertainty of ATN flux on the oscillation parameter experiments, we conduct a fast simulation based on nuSQuIDS~\cite{nuSQuIDS} to investigate the impact of ATN flux uncertainties on the precision measurement of ATN mixing parameters.
We perform a sensitivity analysis of $\theta_{23}$ in a liquid scintillator detector with an exposure of 1500~$\text{kt}\cdot\text{yr}$ as a demonstration.
A fast simulation dataset is generated by folding neutrino cross sections and detector energy resolutions into the HKKM flux model~\cite{Honda:2015fha}. 
The oscillations are computed using nuSQuIDs~\cite{nuSQuIDS}, with $\theta_{23}$ varied and remaining mixing parameters fixed to the PDG values~\cite{ParticleDataGroup:2024cfk}.
The simulated neutrino energy spans the range $-1 < \log_{10}(E/\text{GeV}) < 2$, divided into 100 bins of equal width in $\log_{10}(E/\text{GeV})$.
Due to the limited angular resolution of liquid scintillator detectors, the neutrino direction is integrated out.
For the sake of simplicity, flavor misidentification effects are neglected, and the energy resolution is modeled as~\cite{SFGeHHIF2025}
\begin{equation}
\begin{gathered}
    \frac{\sigma_E^{\nu_\mu}}{\text{GeV}} = 0.011 + 0.043\sqrt{\frac{E}{\text{GeV}}}~, \\
    \frac{\sigma_E^{\bar{\nu}_\mu}}{\text{GeV}} = 0.006 + 0.049\sqrt{\frac{E}{\text{GeV}}}~.
\end{gathered}
\end{equation}
Then the spectra for $\nu_\mu$ and $\bar{\nu}_\mu$ are combined to generate the fast simulation dataset.

Our sensitivity analysis utilizes an Asimov dataset from the fast simulation with an exposure of 1500~$\text{kt}\cdot\text{yr}$ and an assumed true oscillation parameter $\theta_{23}^{\text{truth}}$. The Asimov dataset is normalized to a total of $1.39 \times 10^5$ events, based on a predicted $\nu_\mu + \bar{\nu}_\mu$ rate of 34.7 events per $\text{kt}\cdot\text{yr}$~\cite{Super-Kamiokande:2015qek,JUNO:2021tll}. The normalization defines the statistical uncertainty per bin. Systematic uncertainties are included in the analysis via a $\chi^2$ statistic,
\begin{equation}
\chi^2\left(\theta_{23}\right) = \sum_{k=1}^{n_{\text{bins}}} \frac{\left(n_k - \mu_k(\theta_{23})\right)^2}{(\sigma_{n_k})^2 + (\sigma^{\text{flux}}_k)^2 + (\sigma^{\text{xsec}}_k)^2}~,
\end{equation}
where $n_{\text{bins}} = 100$ denotes the number of energy bins, $n_k = \mu_k(\theta_{23}^{\text{truth}})$ is the Asimov event count in bin $k$, and $\mu_k(\theta_{23})$ is the predicted event count under test values of the oscillation parameters.
The uncertainties include the statistical uncertainty $\sigma_{n_k} = \sqrt{n_k}$, the flux shape uncertainty $\sigma^{\text{flux}}k = f_{\text{flux}} \mu_k$, and the cross-section uncertainty $\sigma^{\text{xsec}}k = f_{\text{xsec}} \mu_k$.
The fractional uncertainties are taken as $f_{\text{flux}} = 10\%, 2\%$ and $f_{\text{xsec}} = 15\%, 2\%$, respectively.

\begin{figure}[htpb]
    \centering
    \begin{subfigure}{\linewidth}
        \centering
        \includegraphics[width = \linewidth]{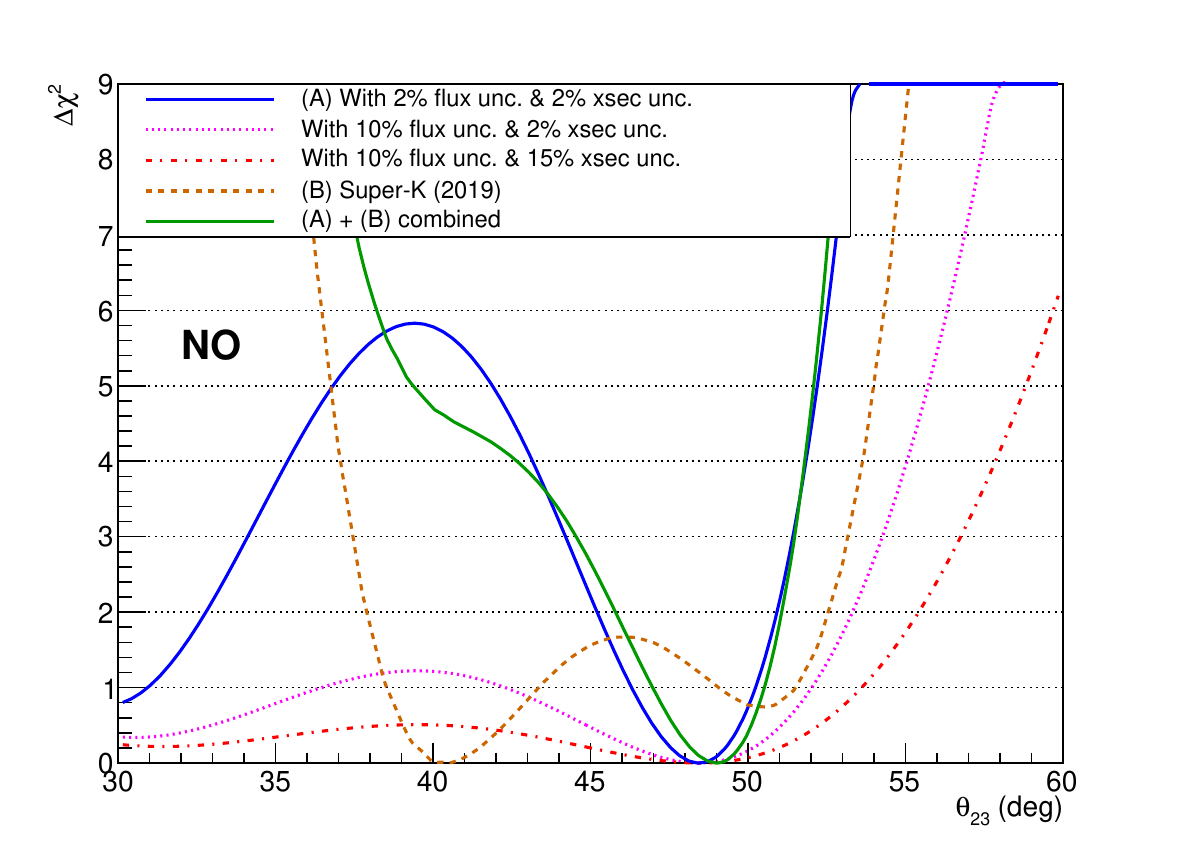}
        \caption{}
        \label{fig:theta23_coincident_no}
    \end{subfigure}
    
    \begin{subfigure}{\linewidth}
        \centering
        \includegraphics[width = \linewidth]{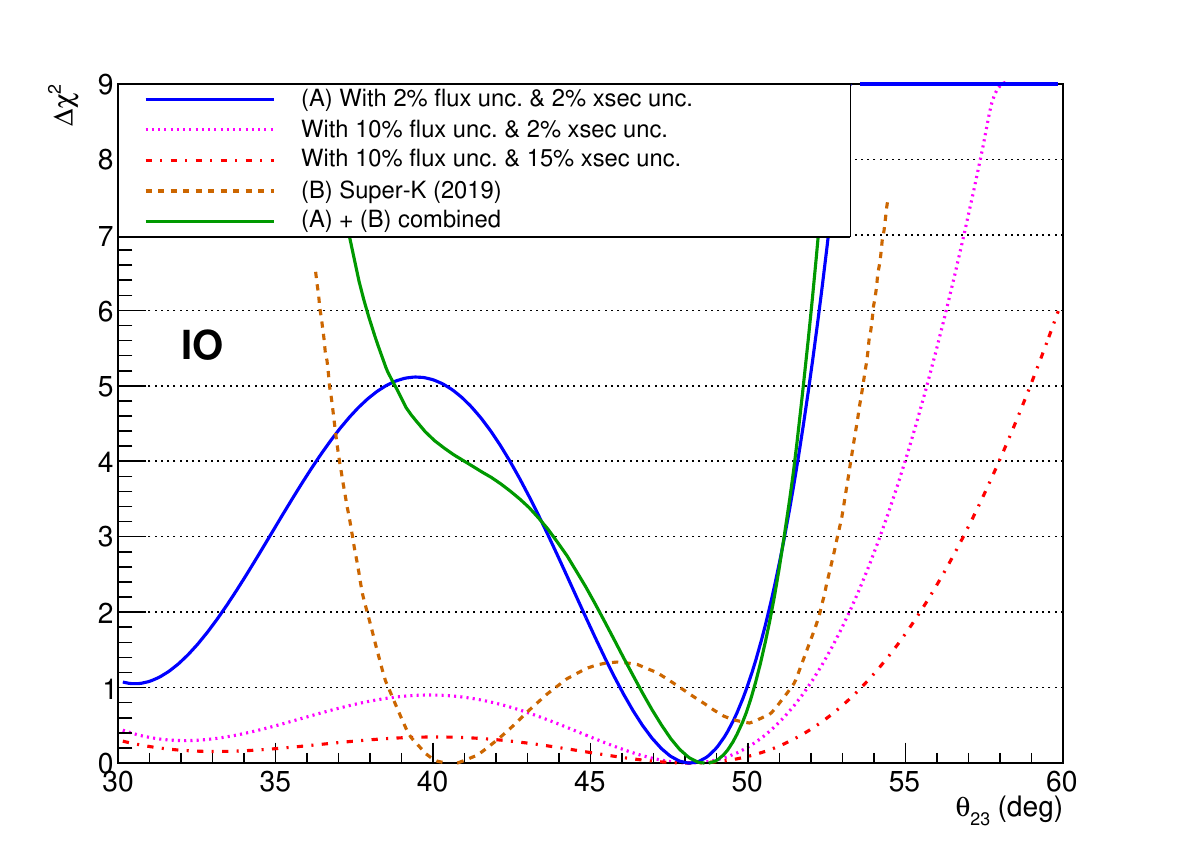}
        \caption{}
        \label{fig:theta23_coincident_io}
    \end{subfigure}
    \caption{\justifying Precision measurement of the mixing angle $\theta_{23}$ with different flux and cross-section uncertainties at ATN experiment utilizing liquid scintillator (LS) with an exposure of 1500~$\text{kt}\cdot\text{yr}$, compared with Super-K measurements~\cite{Super-Kamiokande:2019gzr}, and combined results in normal (NO) and inverse (IO) mass ordering, respectively.}
    \label{fig:ATN_diff_uncertainty}
\end{figure}

The result is presented in~\cref{fig:ATN_diff_uncertainty}.
This indicates that reducing the uncertainty of the ATN flux model significantly improves the sensitivity to $\theta_{23}$.
It offers a unique contribution to the global fit of neutrino mixing parameters, helping to break the octant ambiguity.
Therefore, it is necessary to measure properties of CR muons, including energy spectra, arrival direction, and spin polarization, to further constrain the ATN flux model.

According to the previous discussion, it is necessary to constrain the ATN flux uncertainty, especially the uncertainty of $Y_{p \to \nu}(Y_{A \to \nu})$.
We find that the CR muon polarization $\pmu$ is sensitive to the $K$-$\pi$ ratio, which means that $\pmu$ is a good variable to enhance the constraint on $Y_{i \to \nu}$.
$\pmu$ is defined as the projection of spin angular momentum on the direction of motion, $\pmu = \vec{\sigma} \cdot \hat{p}$.
If muons are produced by two-body decays or $K_L^0$ decays, the spin polarization in the center-of-mass frame should be $\pm 1$ or uniformly distributed between $-1$ and $+1$, respectively.
Since the CR muon polarization is typically measured in the laboratory frame, $\pmu$ in experimental measurement is the polarization after the Lorentz transformation from the center-of-mass frame to the laboratory frame.
To determine the polarization of a muon, one needs to know both its momentum and spin vector.
We begin with a pion traveling along the $x$-axis. 
The four-momenta of the muon and neutrino resulting from the pion decay in the laboratory frame can be written as
\begin{equation}
    \begin{aligned}        
        p^\mu (\mu) &= 
        \frac{M}{2} \left(
            \begin{matrix}
                \cos \alpha \cdot R_m^2 R_p + (1 + R_p^2)^{3/2}\\
                \frac{1 - R_m^2}{1 - R_p (1 + R_p^2)} \cos \alpha \sqrt{1 + R_p^2} \\
                \sin \alpha (1 - R_m^2) \\
                0
            \end{matrix}
            \right),
            \\
            \\
            p^\mu (\nu) &= 
            \frac{M}{2}
            \left(
                \begin{matrix}
                    (1 - R_m^2)(\sqrt{1 + R_p^2} + \cos \alpha R_p)\\
                    -(1 - R_m^2)( \cos \alpha \sqrt{1 + R_p^2} + R_p)\\
                    - \sin \alpha (1 - R_m^2)\\
                    0
                \end{matrix}
                \right),
    \end{aligned}
\end{equation}
where $p^\mu(\nu)$ is the four-momentum of the neutrino, $M$ is the mass of the parent pion, $R_m = m_\mu / M$ is the muon-to-pion mass ratio, $R_p = p_x / M$ is the ratio of the pion momentum to its mass, and $\alpha$ is the angle between the muon's direction and the Lorentz boost axis.
Since neutrinos are exclusively left-handed, their spin vectors are antiparallel to their momentum.
By conservation of angular momentum in the pion rest frame, the muon’s spin direction is opposite to that of the neutrino. 
Applying the Lorentz transformation to both momentum and spin, the muon polarization vector in the laboratory frame, $\pmu$, can be expressed analytically in terms of $R_p$ and $R_m$:
\begin{equation}
    \begin{aligned}
        \pmu = \frac{\cos \alpha \sqrt{1 + R_p^2} + R_p + g(\alpha, R_p, R_m)}{h(\alpha, R_p) \sqrt{1 + g^2(\alpha, R_p, R_m)}}\, ,
    \end{aligned}
    \label{eq:polarization_laboratory_sup}
\end{equation}
where
\begin{equation}    
    \begin{aligned}
        &g(\alpha, R_p, R_m) = \frac{\cos \alpha \sqrt{1 + R_p^2} + R_p + R_m}{\sqrt{1 + R_m^2} h(\alpha, R_p)}\, ,
        \\
        &h(\alpha, R_p) = \sqrt{2 R_p \cos \alpha \sqrt{1 + R_p^2} + (1 + \cos^2 \alpha) R_p + 2}\, .
    \end{aligned}
\end{equation}

\begin{figure}[htpb]
    \centering
    \includegraphics[width = \linewidth]{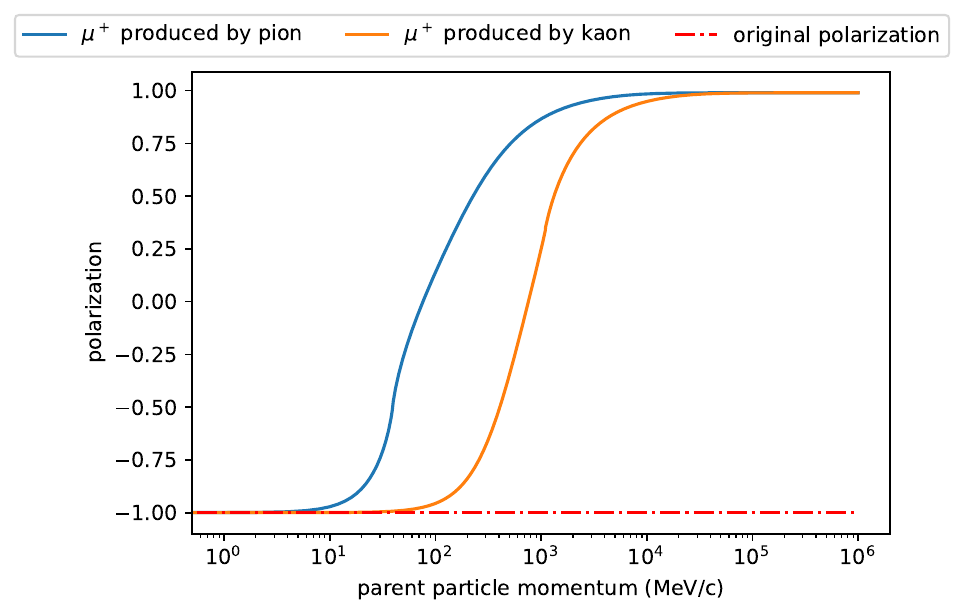}
    \caption{\justifying Polarization of $\mu^+$ produced via two-body decays of pions and kaons as a function of the parent particle momentum. Owing to the larger kaon mass, the resulting muon polarization varies more mildly compared to that from pions. At high parent momentum, the muon polarization approaches $+1$, transitioning from the initial value of $-1$ in the rest frame.}
    \label{fig:Pion_Kaon_polarization}
\end{figure}

According to~\cref{eq:polarization_laboratory_sup}, $\pmu$ in the laboratory frame depends on the mass and the momentum of the parent particle.
The average polarization of muons produced by pions and kaons as a function of the parent particle momentum is shown in~\cref{fig:Pion_Kaon_polarization}, where the average polarization is obtained by integrating over $\alpha$ from $0$ to $2\pi$.
The plot shows a difference in average polarization between muons produced by pions and kaons, even if the parent particles have the same momentum.
Therefore, measuring $\pmu$ provides a unique handle to determine the relative contributions of pions and kaons in cosmic ray muon production, which in turn constrains the {\kpi} in ATN production.

\begin{figure}[htpb]
    \centering
    \includegraphics[width=0.7 \linewidth]{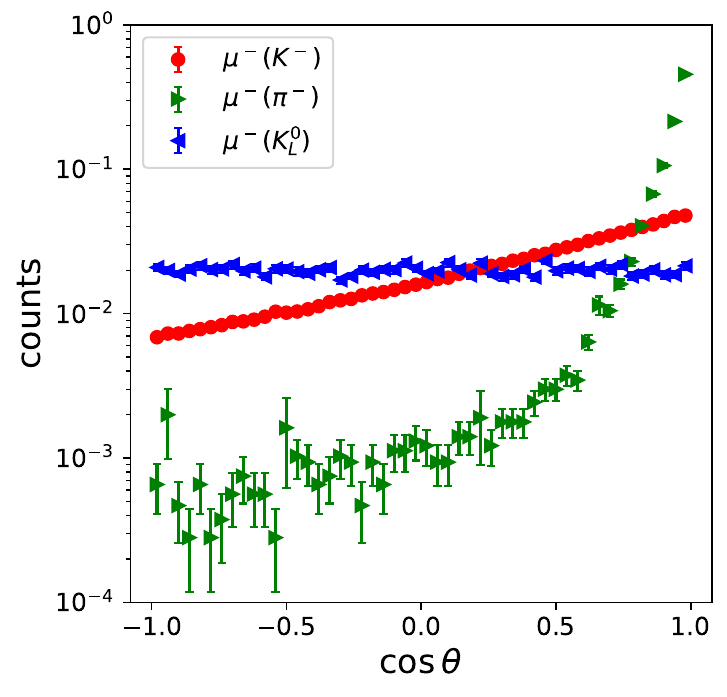}
    \caption{\justifying Normalized $\pmu$ distributions of cosmic-ray muons produced by different parent particles in MusAirS simulations. Red circles, green right-pointing triangles, and blue left-pointing triangles denote polarization for muons from pion decays, kaon, and $K_L^0$ decays, respectively.}
    \label{fig:MusAirSPmu}
\end{figure}

\noindent\textbf{Simulation.}
To study the CR muon polarization $\pmu$ in the experiment, we concentrate on the muon polarization in MusAirS.
MusAirS can effectively simulate the depolarization process when cosmic ray muons penetrate the atmosphere in a broad energy range $[10^2, 10^5]~\mathrm{MeV}$.
To align closely with experimental conditions, we focus on cosmic ray muons with kinetic energy between $100~\mathrm{MeV}$ and $500~\mathrm{MeV}$ and zenith angles less than $\pi / 3$.
As shown in~\cref{fig:MusAirSPmu}, different parent particles of CR muons result in distinct distributions of $\pmu$.
Due to the mass difference, the polarization distribution of muons produced by pion decays is flatter than that of muons produced by kaons.
As a result, the polarization distribution of pion-produced and kaon-produced muons exhibits a notable difference.
The polarization of cosmic-ray muons affects the angular distribution of Michel electrons from their decays. 
We define $\Nr$ as the distribution of Michel electrons with respect to the muon momentum direction, where $\theta_r$ is the angle between the muon and electron momenta. This distribution $\Nr$ can be written as a function of the muon polarization vector $\pmu$,
\begin{equation}
    \label{eq:Michel_distribution}
    \begin{aligned}
        \frac{\mathrm{d}\Nr}{\mathrm{d} \phi \mathrm{d} \cos \theta_r} \propto  
        \left(1 + \frac{1}{3} \pmu \cos \theta_r \right)
    \end{aligned}.
\end{equation}

\begin{figure}[htpb]
    \centering
    \includegraphics[width=0.7 \linewidth]{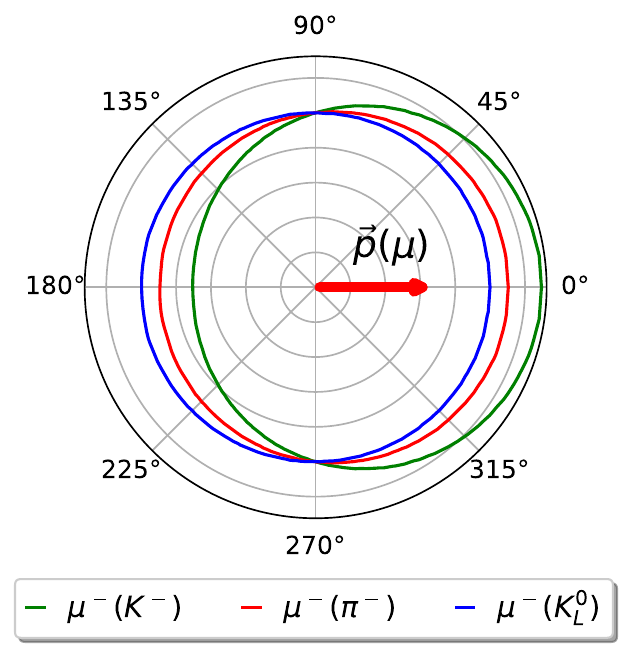}
    \caption{\justifying
    $\Nr$ of muons produced by different parent particles.
    The red angle indicates the direction of the muon momentum.
    The green, red, and blue lines indicate $\Nr$ for muons produced by kaon, pion, and $K_L^0$, respectively.
    }
    \label{fig:MusAirSMichelElectron}
\end{figure}

\begin{figure}[htpb]
    \centering
    \includegraphics[width = \linewidth]{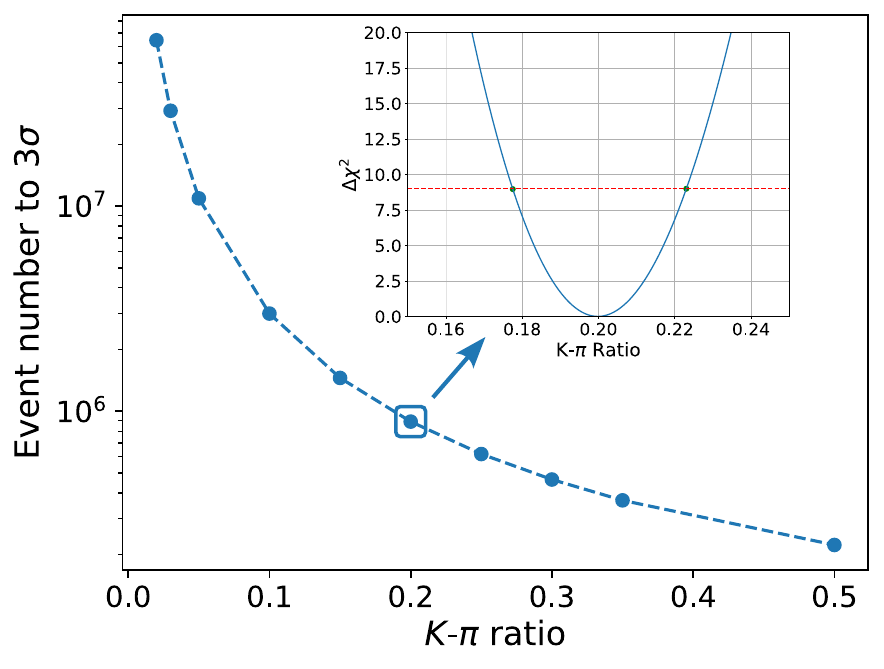}
    \caption{\justifying Estimation of the Data Required for Measuring Cosmic-Ray Muon Polarization. The main panel displays the amount of data needed to reject the null hypothesis of a pure pion parent population. The horizontal axis represents the assumed input ratio $ R_{K\text{-}\pi} $, while the vertical axis shows the corresponding $\chi^2$ quantile (denoted as $1-p$). 
    The inset illustrates the expected experimental sensitivity to $R_{K\text{-}\pi}$ for a true value of $0.200$. This is based on $3 \times 10^5$ Michel-electron events, corresponding to 1 year of data collection with the CRmuSR array.}
    \label{fig:sensitivity_extimation}
\end{figure}

The simulated distributions of $\Nr$ under different scenarios are shown in~\cref{fig:MusAirSMichelElectron}.
Michel electrons from kaon-decay muons exhibit a strong alignment with the muon momentum, while those from pion-decay muons are more isotropic but still retain a moderate directional preference.
In contrast, muons originating from $K_L^0$ decays yield an almost isotropic Michel electron distribution.
These distinct spatial patterns of Michel electrons, associated with muons from different parent particles, provide a way to disentangle the relative contributions of pions, kaons, and $K_L^0$, commonly denoted as {\kpi} ($R_{K\text{-}\pi}$).
It follows that the observable $\Nr$ can be expressed as a superposition of three components, each corresponding to muons from a given parent type, as illustrated in~\cref{fig:MusAirSMichelElectron}.
By performing a spectral analysis of $\Nr$, the value of {\kpi} can be extracted within a specified muon kinetic energy range.
To demonstrate the feasibility of the proposed method, we evaluate it in two steps. 
In the first step, we estimate the number of Michel electron events required to exclude the pure-pion hypothesis at a given significance level. 
A $\chi^2$ test is performed by comparing the measured $\Nr$ distribution with the expectation under the null hypothesis, which assumes that all muons originate from pion decays. 
For simplicity, it is assumed that the detector can identify the Michel electron charge and detect all cosmic-ray muons with kinetic energies between $100$ and $500~\mathrm{MeV}$ and zenith angles in the range $[0, \pi/6]$ with uniform efficiency. 
The spatial distributions of Michel electrons corresponding to different $R_{K\text{-}\pi}$ values are obtained from Monte Carlo simulations, and the $p$-values from the $\chi^2$ tests are converted to Gaussian quantiles. 
When the quantile exceeds $3\sigma$, the pure-pion hypothesis is excluded, indicating a significant kaon contribution. 
The required event number as a function of $R_{K\text{-}\pi}$ is shown in~\cref{fig:sensitivity_extimation}. 
As $R_{K\text{-}\pi}$ increases, fewer events are needed to reject the null hypothesis; however, for $R_{K\text{-}\pi}<0.5$, more than $10^6$ Michel electron events are required, corresponding to over $600~\mathrm{m}^2 \cdot \mathrm{days}$ of exposure at sea level.
Due to seasonal variation in atmospheric neutrino flux, it is recommended to collect data simultaneously and ensure sufficient statistics by deploying a detector array.

In the second step, we evaluate the sensitivity achievable with a well-performing CRmuSR array. 
Assuming an effective area of $\mathcal{O}(10)\mathrm{m^2}$, the expected Michel electron statistics in one year amount to $3 \times 10^5$ events, corresponding to an event rate of $3 \times 10^4\mathrm{m^{-2}\cdot yr^{-1}}$. 
To validate the analysis, the fast simulation sample is compared with an Asimov dataset generated under the same {\kpi} assumption, with $R_{K\text{-}\pi} = 0.20$ fixed in accordance with~\cite{Super-Kamiokande:2024rwz}.
In this sensitivity study, the detector is assumed to identify the muon charge and fully reconstruct the Michel electron angular distribution, while systematic uncertainties are neglected under idealized conditions. 
The test statistic is defined as
\begin{equation}
    \chi^2 (R_{K\text{-}\pi}) = \sum_{k=1}^{n_\text{bins}} \frac{(n_k - \mu_k (r_{K \pi}))^2}{\sigma_{k}^2},
\end{equation}
where $n_\text{bins} = 100$ is the number of $\cos\theta_Z$ bins, $n_k = \mu_k(0.200)$ is the Asimov event count in bin $k$, $\mu_k(R_{K\text{-}\pi})$ is the expected count under a test value of $R_{K\text{-}\pi}$, and $\sigma_k = \sqrt{n_k}$ is the statistical uncertainty. 
The $\chi^2$ is minimized to determine the best-fit value of $R_{K\text{-}\pi}$, with the $1\sigma$ confidence interval defined by $\Delta \chi^2 < 1$. 
As shown in~\cref{fig:sensitivity_extimation}, the expected $68\%$ confidence interval is $[0.176, 0.224]$, indicating that a CRmuSR array of $\mathcal{O}(10)~\mathrm{m^2}$ can achieve a precision measurement of $R_{K\text{-}\pi}$ within one year of operation under optimistic assumptions.

\begin{figure}[htpb]
    \includegraphics[width = \linewidth]{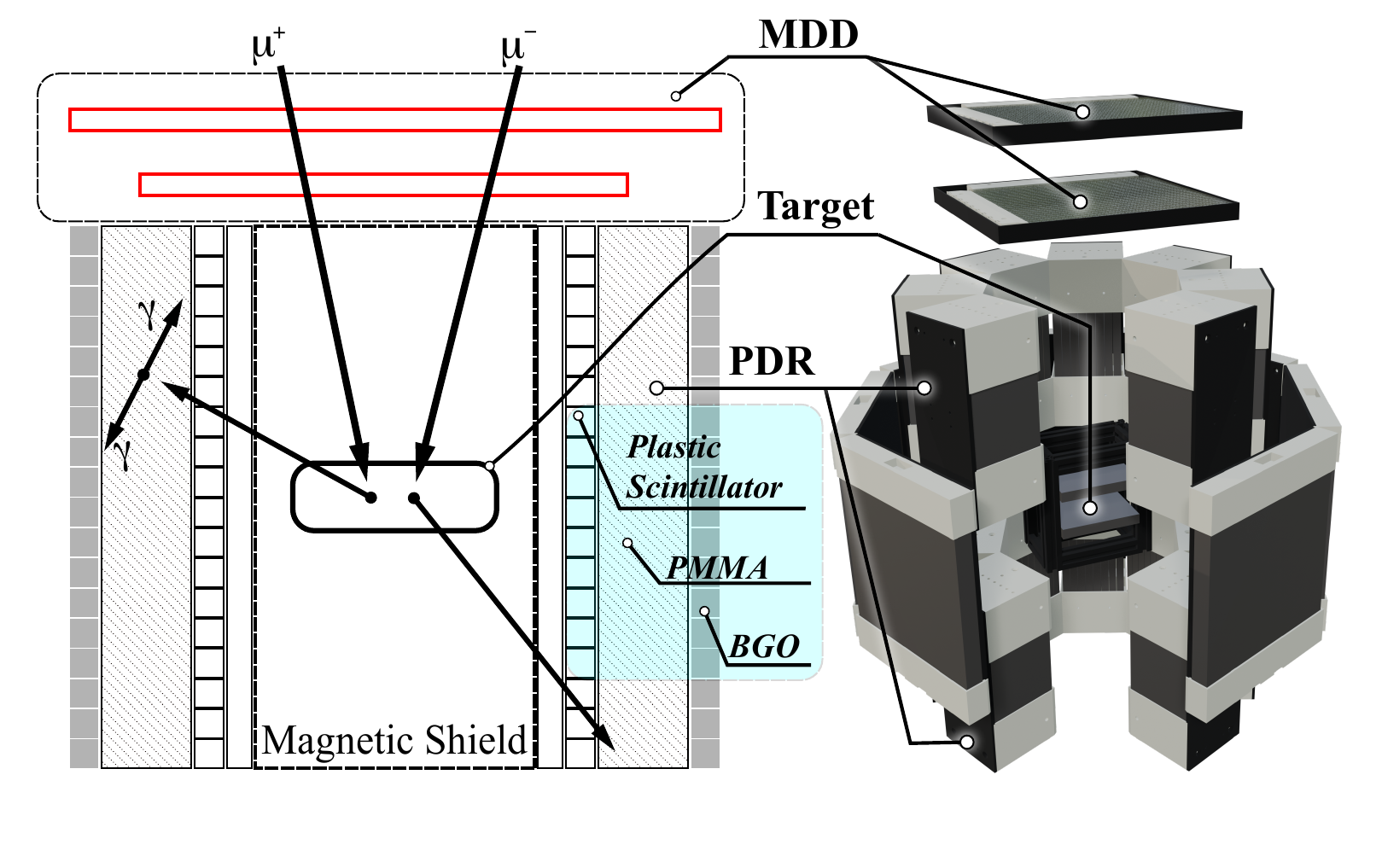}
    \caption{\justifying
    Design of CRmuSR.
    The left panel is the cross-sectional view of CRmuSR.
    The top dashed rectangle is the MDD, composed of multiple layers of position-sensitive detectors.
    In the center is the cosmic-ray muon stopping target.
    Surrounding the target is the PDR.
    The PDR consists of a plastic scintillator array, an absorber to stop the Michel electrons and positrons, and a ring of BGO crystals to detect the $\mathrm{\gamma}$ produced by positron annihilation.
    The right panel displays the design diagram of the prototype, showing the PDR in green, the MMD in red, and the target in gray.}
    \label{fig:CRmuSRDesign}
\end{figure}

\noindent\textbf{Cosmic-Ray muon Spin polarization detectoR (CRmuSR).}
CRmuSR is a modular detector system designed to measure cosmic-ray muon polarization $\pmu$ via the angular distribution of Michel electrons.
Its array-based configuration enables parallel deployment of multiple modules, significantly reducing data-taking time and allowing studies of seasonal variations in atmospheric neutrinos (ATNs).
As illustrated in~\cref{fig:CRmuSRDesign}(a), each module consists of a Momentum Direction Detector (MDD), a muon-stopping target, and a surrounding Positron/Electron Detector Ring (PDR).
The MDD reconstructs the incoming muon’s momentum direction using multiple layers of position-sensitive detectors, and the PDR consists of plastic scintillators and BGO detectors to detect Michel electrons and their emission angles.
Together, the MDD and PDR allow the reconstruction of the observable $\Nr$, defined as the angular distribution of Michel electrons relative to the muon momentum.
A prototype module has been constructed and tested, as shown in~\cref{fig:CRmuSRDesign}(b).
The MDD achieves a zenith angle resolution better than $2^\circ$, and the partially implemented PDR covers approximately $40\%$ of the solid angle, with angular resolution better than $6^\circ$. 
A copper slab serves as the muon-stopping target to minimize depolarization effects~\cite{PhysRev.112.580}.
In a 1000-hour data run, cosmic-ray muon events were identified via MDD coincidences within a $5~\mathrm{ns}$ window, and associated Michel electrons were detected within $20~\mathrm{\mu s}$ using the PDR.
Using electron counts from the upper and lower PDR rings, we measured the up-down asymmetry of Michel electrons as $A_\text{exp} = -0.008 \pm 0.044$, which does not show a statistically significant deviation from the theoretical expectation $A_\text{th} = 0.019$, based on the muon charge ratio from MINOS~\cite{MINOS_charge_ratio} and the polarization prediction by Honda~\cite{Super-Kamiokande:2024rwz}.

\noindent\textbf{Conclusion and outlook.}
We have developed a novel experimental setup to probe the polarization of low-energy cosmic-ray muons, enabling the reconstruction of the parent particle composition relevant to atmospheric neutrino studies. 
Simulations indicate that the angular distribution of Michel electrons provides a viable observable for determining the polarization profile of cosmic-ray muons.

To demonstrate feasibility, we constructed a prototype detector system, CRmuSR, and performed initial measurements of Michel electron asymmetry. Its modular design and use of cost-effective scintillators allow straightforward scaling to large arrays.
A full-scale deployment of CRmuSR with an effective area of $\mathcal{O}(10)~\mathrm{m}^2$ in a year is expected to provide valuable input on atmospheric neutrino fluxes, offering complementary information for next-generation neutrino experiments such as DUNE, Hyper-Kamiokande, and JUNO, and upcoming water-based scintillator detectors~\cite{Land:2020oiz, Theia:2022uyh, Luo_2023}, which may also benefit from external constraints on the parent particle composition. 
It is envisaged that the oscillation analysis of $\mathcal{O}(1)$ GeV-scale atmospheric neutrinos in a liquid scintillator detector with an exposure of 1500~$\text{kt}\cdot\text{yr}$ will break the octant degeneracy and achieve the precision measurement of $\theta_{23}$ with the uncertainty smaller than $5^\circ$ at 3$\sigma$ confidence level irrespective of the mass ordering.

\noindent\textbf{Acknowledgments.}
This project was supported in part by National Natural Science Foundation of China under Grant Nos. 12347105 and 12075326, the Natural Science Foundation of Guangzhou under Grant No. 2024A04J6243 and Fundamental Research Funds for the Central Universities
(23xkjc017) in Sun Yat-sen University. J.T. is grateful to Southern Center for Nuclear-Science Theory (SCNT) at Institute of Modern Physics in Chinese Academy of Sciences for hospitality. The simulation was conducted by a strong support of computing resources from the National Supercomputer Center in Guangzhou.

\bibliography{bib}

\end{document}